\documentclass[aps,prc,reprint,groupedaddress]{revtex4-2}
\usepackage{lineno,hyperref, amssymb, amsthm, graphicx, amsmath, color, multirow}

\begin{document}


\title{Decay of superheavy nuclei based on the random forest algorithm}


\author{Boshuai Cai}
\affiliation{Sino-French Institute of Nuclear Engineering and Technology, Sun Yat-sen University, Zhuhai, 519082}

\author{Cenxi Yuan}
\email[]{yuancx@mail.sysu.edu.cn}
\affiliation{Sino-French Institute of Nuclear Engineering and Technology, Sun Yat-sen University, Zhuhai, 519082}


\date{\today}

\begin{abstract}

How nuclides decay in the superheavy region is key information for investigating new elements beyond oganesson and the island of stability. The Random Forest algorithm is applied to study the competition between different decay modes in the superheavy region, including $\alpha$ decay, $\beta^-$ decay, $\beta^+$ decay, electron capture and spontaneous fission. The observed half-lives and dominant decay mode are well reproduced. The dominant decay mode of 96.9 \% nuclei beyond $^{212}$Po is correctly described.  $\alpha$ decay is predicted to be the dominant decay mode for isotopes in new elements $Z = 119 - 122$, except for spontaneous fission in some even-even ones because of the odd-even staggering effect. The predicted half-lives show the existence of a long-lived spontaneous fission island at the southwest of $^{298}$Fl caused by the competition of nuclear deformation and Coulomb repulsion. More understanding of spontaneous fission, especially beyond $^{286}$Fl, is crucial to search for new elements and the island of stability.

\end{abstract}


\maketitle

\section{Introduction}

The limit of nuclear landscape \cite{nazarewicz2018, erler2012} is always an intriguing subject. Exotic properties of nuclei are found around the boundary of nuclear limits, e.g., the shell evolution \cite{otsuka2020, otsuka2010, ozawa2000, smirnova2010}, the 4$n$ resonant state \cite{duer2022, li2019}, the 4$p$ unbound state \cite{jin2021}, etc. The discovery of new elements (nuclides) generally faces three problems: production, separation and identification \cite{hofmann2000}. Because the nucleus is unstable, generally with a rather short half-life, one has to utilize some probes. One of the most direct is the decay mode \cite{hofmann2000, giuliani2019}, using the decay products as the signal of existence. Thus, it is important to investigate and predict the dominant decay mode of those unknown nuclides. The nuclear binding energy and the half-life are key data for understanding the decay mode of an atomic nucleus. The former measures the stability of nuclides through energy criteria, and the latter describes the possibility of decay.

Both microscopic and macroscopic methods have been used to study the nuclear binding energy and the partial half-life of each decay channel. The microscopic theory starts from the nucleon-nucleon interaction, either realistic or phenomenological. The macroscopic theory uses selected variables with physical considerations to construct semi-empirical formulas and fit the experimental data, with a risk of overfitting and inappropriate parameters. Besides, the exotic nuclei may deviate far from the general fitting as outliers. Decreasing the deviation between theoretical predictions and observables is always a critical issue.

With computing and storage power advancement, machine learning algorithms have become available and helpful in many fields with various successes \cite{mehta2019}. As summarized in a recent colloquium, estimating the residuals of nuclear properties through the machine learning algorithms is a powerful strategy \cite{boehnlein2022}. The neural network was used to compensate for the residuals of nuclear masses \cite{utama2016mass, niu2018, niu2019} and nuclear charge radii \cite{utama2016, wu2020, dong2022} with structure optimization and careful choice of the inputted parameters with definite physical meanings. The applicability of the Decision Tree (DT) was verified by training and testing with residuals of binding energy in 2020 \cite{carnini2020}. However, the Random Forest (RF) \cite{breiman2001}, developing from the DT algorithm, has been tested for neither nuclear mass nor partial half-life of a specific decay channel, of which the semi-empirical formulas have suggested several major components but with residuals. The machine learning algorithms can include the possible features to make a training for the residuals, while the RF, with bootstrap sampling, can not only avoid the overfitting but also take into account the correlation between data combinations and several features, which increases the robustness and is conducive to the extrapolation.

Until now, none of work has investigated the competition between different decay modes by the partial half-lives estimated by the machine learning algorithm. The present work applies the RF machine learning algorithm to study the major decay mode of heavy and superheavy nuclei. The competition of $\alpha$ decay, $\beta$ decay and spontaneous fission (SF) of new elements and the possible long-lived island are discussed in the superheavy region.

\section{Method}

The present work concentrates on the region of $Z \geqslant 84$ and $N \geqslant 128$. The partial half-lives of $\alpha$ decay, $\beta^-$ decay, $\beta^{+}$ decay, electron capture (EC), and SF are calculated by the semi-empirical formulas and then the residuals of each formula are trained by the RF algorithm respectively. The minimum partial half-life of one mode corresponds to the dominant decay mode.

\subsection{Decay Half-life Formulas}
The universal decay law (UDL) \cite{qi2009,qi2009prl},
\begin{equation}
\begin{aligned}
\label{eq:udl}
\text{log}_{10}T_{1/2, \alpha} =&aZ_\alpha (Z-Z_\alpha)\sqrt{\mu/Q_\alpha}\\
 &+ b\sqrt{\mu Z_\alpha (Z-Z_\alpha)(A_\alpha^{1/3}+(A-A_\alpha)^{1/3})}\\
&+ c,\\
\end{aligned}
\end{equation}
is used to fit the $\alpha$ decay half-life. $Z_\alpha$, $A_\alpha$, $Q_\alpha$ and $\mu=A_\alpha (A-A_\alpha)/A$ denote the proton number, the mass number of $\alpha$ particle, the $\alpha$ decay energy and the reduced mass, respectively. The channel is supposed to be from the ground state to the ground state.

 \begin{table*}[htbp]
 \caption{\label{tab:T}The coefficients and the corresponding RMSE of UDL, SF3 and Eqs. (\ref{eq:ren}, \ref{eq:xu}, \ref{eq:santhosh}, \ref{eq:soylu}, \ref{eq:beta}) when fitted to the nuclei with $Z\geqslant84$ \& $N \geqslant 128$. The RMSEs of RF trained UDL, SF3 and Eq. (\ref{eq:beta}) are listed in the last two columns. WS4 and UNEDF0 in the subscript denotes that the source of the unmeasured energies.}
\begin{ruledtabular}
\begin{center}
\begin{tabular}{cccccccccc}
		&$a$ (log$_{10}B_{\text{GT}}$) 	& $b$ 	& $c$ 	& $d$ 	& $e$ 	& $f$	& RMSE	& RMSE$_{\rm{RF,WS4}}$ & RMSE$_{\rm{RF,UNEDF0}}$		\\
\hline
UDL			&0.407	&-0.382	&-23.896	&-		&-		&-		&0.883	&0.598 & 0.669				\\
SF3 ($Z<104$)	&-1.129	&-6997.113&79.803	&-		&-		&-		&3.070	&1.195 & 1.195				\\
SF3 ($Z\geqslant 104$)	&-1.363	&-13272.729&113.415	&-		&-		&-		&1.267	&0.825 & 0.825				\\
SF$_{\text{ren}}$&-1002.490&35.340	&-0.332	&-0.747	&31.565	&-		&2.875	&- &					\\
SF$_{\text{xu}}$&-1.031	&0.145	&6.28E-07	&-0.00986	&-1.135	&572.558	&3.313	&- & 					\\
SF$_{\text{santhosh}}$&-37.871&0.432	&3325.452	&-8270.979	&496.073	&-	&2.857	&- & 					\\
SF$_{\text{soylu}}$&-19.408	&180.953	&-0.491	&-0.807	&3.07E-06	&-1851.543&3.181	&- & 					\\
Eq. (\ref{eq:beta})$_{\beta^+}$		&1.378	&-		&-		&-		&-		&-		&1.957	&0.439 & 0.437				\\
Eq. (\ref{eq:beta})$_{\beta^-}$		&-1.819	&-		&-		&-		&-		&-		&1.451	&0.656 & 0.667				\\
Eq. (\ref{eq:beta})$_{\text{EC}}$		&-2.112	&-		&-		&-		&-		&-		&2.360	&0.971 & 0.996				\\
\end{tabular}
\end{center}
\end{ruledtabular}
 \end{table*}

As to SF, a three-parameter formula (noted as SF3),
\begin{equation}
\label{eq:sf}
\log_{10}T_{\mathrm{SF}} = a\frac{(Z-\nu)^2}{(1-\kappa I^2)A}+\frac{b}{A}+c, 
\end{equation}
is proposed based on several existing formulas \cite{ren2005, xu2008, santhosh2010, soylu2019, bao2015, santhosh2021}, where $\nu$ presents the blocking effect from unpaired nucleons, takes 0 for even-even nuclei and 2 for other nuclei \cite{ren2005}, $\kappa$ takes 2.6 \cite{Royer2000, bao2015}, $I=\frac{N-Z}{A}$, and $a$, $b$, and $c$ are fitting coefficients. Eq. (\ref{eq:sf}) is particularly fitted to nuclei with $Z<104$ and the rest because of a systematic difference as shown in TABLE \ref{tab:T}. $T_{1/2,\text{SF}}$ of nuclei with $Z<104$ increases largely with the decrease of $Z$ since the Coulomb repulsion decreases. The rather long $T_\text{SF}$ ($> 10^8$ s) of some nuclei in this region cannot be universally described at present and are not taken to fit Eq.(\ref{eq:sf}) because the competition of such SF is very weak comparing to other decay modes. This formula avoids the divergence of other SF formulas in Refs. \cite{ren2005, xu2008, santhosh2010, soylu2019} during extrapolation, which write as:
\begin{equation}
\begin{aligned}
\label{eq:ren}
\text{log}_{10}T_{1/2, \text{SF}, \text{ren}}=&a\frac{Z-90-\nu}{A}+b\frac{(Z-90-\nu)^2}{A}\\
&+c\frac{(Z-90-\nu)^3}{A}\\
&+d\frac{(Z-90-\nu)(N-Z-52)^2}{A} + e,
\end{aligned}
\end{equation}
\begin{equation}
\begin{aligned}
\label{eq:xu}
\text{log}_{10}T_{1/2, \text{SF}, \text{xu}}=&aA+bZ^2+cZ^4+d(N-Z)^2\\
 &+eZ^2A^{-1/3}+f,
\end{aligned} 
\end{equation}
\begin{equation}
\begin{aligned}
\label{eq:santhosh}
\text{log}_{10}T_{1/2, \text{SF}, \text{santhosh}}=&a\frac{Z^2}{A}+b(\frac{Z^2}{A})^2\\
&+c\frac{N-Z}{A}+d(\frac{N-Z}{A})^2+e,
\end{aligned} 
\end{equation}
\begin{equation}
\begin{aligned}
\label{eq:soylu}
\text{log}_{10}T_{1/2, \text{SF}, \text{soylu}}&=aA+bA^{2/3}+cZ(Z-1)A^{-1/3}\\
&+d(N-Z)^2/A+eZ^4+f.
\end{aligned} 
\end{equation}
When the higher order terms in these four formulas enhance their interpolation, the divergence is introduced into the extrapolation, which will be discussed in the last part of Sec. \ref{sec:results}.

The $\beta$ decay half-life is estimated by the formula in Refs. \cite{suhonen2007, gao2022}.  Assuming that the ground state $\beta$ decay is one effective Gamow-Teller (GT) transition, the partial half-life is expressed as 
\begin{equation} 
\begin{aligned}
\label{eq:beta}
\text{log}_{10}T_{1/2,\beta}=\text{log}_{10}\kappa_1-\text{log}_{10}f_0-\text{log}_{10}B_{\text{GT}},
\end{aligned}
\end{equation}
where $\kappa_1=\frac{2\pi^3\hbar^7\ln2}{m_e^5c^4G_\text{F}^2}=6147$ s, $f_0$ is the phase-space factor and $B_{\text{GT}}$ is the GT reduced transition probability \cite{suhonen2007}. For EC, the phase-space factor is deduced as 
\begin{equation} 
\begin{aligned}
f_0^{\text{EC}} \approx 2\pi(\frac{Z}{137})^3(1-\frac{1}{2}(\frac{Z}{137})^2+E_0)^2,
\end{aligned}
\end{equation}
while for $\beta^{\pm}$ decay, it is 
\begin{equation} 
\begin{aligned}
f_0^{\beta^{\pm}} \approx \frac{\mp(E_0^5-10E_0^2+15E_0-6)2\pi (Z\mp1)/137}{30(1-\exp(\pm2\pi (Z\mp1)/137))},
\end{aligned}
\end{equation}
where $E_0$ is the renormalized $\beta$ decay energy. Because $Q_\beta$ provided by AME2020 \cite{wang2021} is the difference of atomic mass,  the electron mass should be reconsidered:
\begin{equation}
\begin{aligned}
E_{0, \beta^+} &= \frac{Q_{\beta^+}+2m_ec^2}{m_ec^2}\\
E_{0, \beta^-} &= \frac{Q_{\beta^-}+m_ec^2}{m_ec^2}\\
E_{0, \text{EC}} &= \frac{Q_{\text{EC}}-m_ec^2}{m_ec^2}.\\
\end{aligned}
\end{equation}
Finally, the log$_{10}B_{\text{GT}}$ is estimated as the average of log$_{10}(f_0T_{1/2, \beta}/\kappa_1)$. The fitting results are listed in TABLE. \ref{tab:T}

\subsection{Random Forest Method}

RF is an integration of the DT and bootstrap algorithm. DT is a non-parametric supervised learning algorithm. For a dataset consisting of $S$ samples of $I$ features (variables) $\{(\theta_1, ..., \theta_I)_s, s\in [1,S]\}$ and object (observable) $\{y_s, s\in [1,S]\}$, it establishes a binary tree structure which divides the dataset into $L$ subsets based on the values of features, each subset is called as a leaf. Such partition aims at the minimum root-mean-square error (RMSE)
\begin{equation}
\mathrm{RMSE} = \sqrt{\frac{1}{S}\sum_{s=1}^S (y_s - f(\theta_1, ..., \theta_I))^2} 
\end{equation}
 of the whole dataset, assigning each leaf a value. 

Bootstrap is a statistical method with a basic idea of randomly resampling with replacement, through which the possible combination and weighting of data are automatically taken into account \cite{cai2020, cai2022}. Each time a new dataset is obtained, a new DT is trained and used to predict the object of each sample in the whole dataset. Repeating this $M$ times, one obtains a forest of $M$ trees. The final predicted value of the object for a sample is the average of results calculated by all trees in the forest. Since each tree is trained by a part of samples in the dataset, the value for each sample predicted by the forest is an average of interpolation and extrapolation, which decreases the divergence when the calculation is implemented to the unmeasured nuclei. The open source scikit-learn \cite{pedregosa2011} is used for machine learning. The forest is assumed to be composed of $10^{5}$ trees to decrease the dispersion of RMSE in the present work.

\section{Results and Discussion\label{sec:results}}

\begin{figure}[htbp]
\begin{center}
\includegraphics[height=0.45\textwidth]{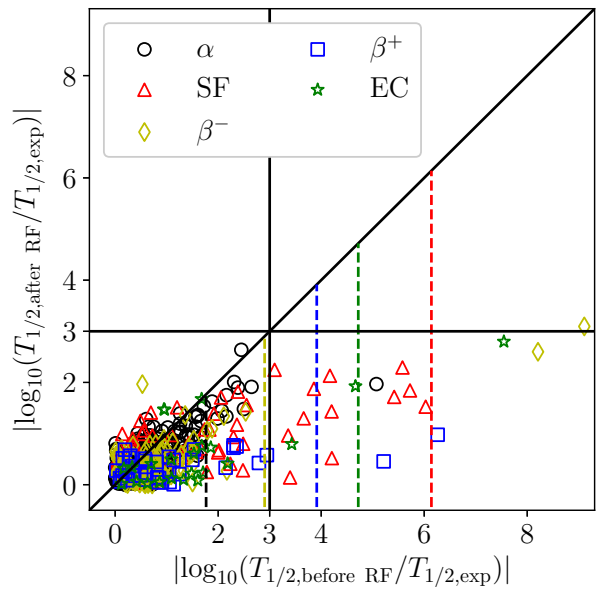}
\caption{\label{fig:outliers} Comparison of residuals of UDL ($\alpha$ decay), SF3 (SF), Eq. (\ref{eq:beta}) ($\beta^-$,  $\beta^+$ and EC) before and after RF training. The dash lines denote two times of RMSE of corresponding formulas.}
\end{center}
\end{figure}

The residuals of the decay formulas of $\alpha$ decay, $\beta^-$ decay, $\beta^+$ decay, and EC are trained by RF with features $Z$, $N$, $A$, the odevity of $Z$ and $N$, and decay energy. Because no decay energy can be defined for SF, the fission barrier (FB) extracted from Ref. \cite{moller2015} is used to replace the decay energy in the features set to consider the deformation effect. The leaves number chooses 11, which is same as that determined for training the binding energy in this region in our previous work \cite{cai2022LDRF}.  FIG. \ref{fig:outliers} compares the residuals of these decay formulas before and after RF training. Two conditions are assumed to determine outliers here. One is locating out of the dash line with corresponding color, which means the scatters deviate two times of RMSE from the experimental $\mathrm{log}_{10}T_{1/2}$. Another is $|\mathrm{log}_{10}(T_{1/2, \mathrm{cal}}/T_{1/2, \mathrm{exp}})|$ value larger than 3, which denotes that the calculation is three times of magnitude far from the experimental value. In this way, one avoids missing (adding) outliers due to the very large (small) RMSE. After training, the biases of outliers of these decay formulas are well reduced,  and the RMSE of formulas decreases (TABLE \ref{tab:T}) as expected. The condition of outlier is not too strict because our aim is not to decrease the RMSE as small as possible but go to an appropriate scale where the dominant decay mode can be described. The present work chooses the same features and the same leaves number of RF to train the residuals of different decay formulas, which avoids the overfitting in seeking an extreme small RMSE.

\begin{figure*}[htbp]
\begin{center}
\includegraphics[height=0.28\textwidth]{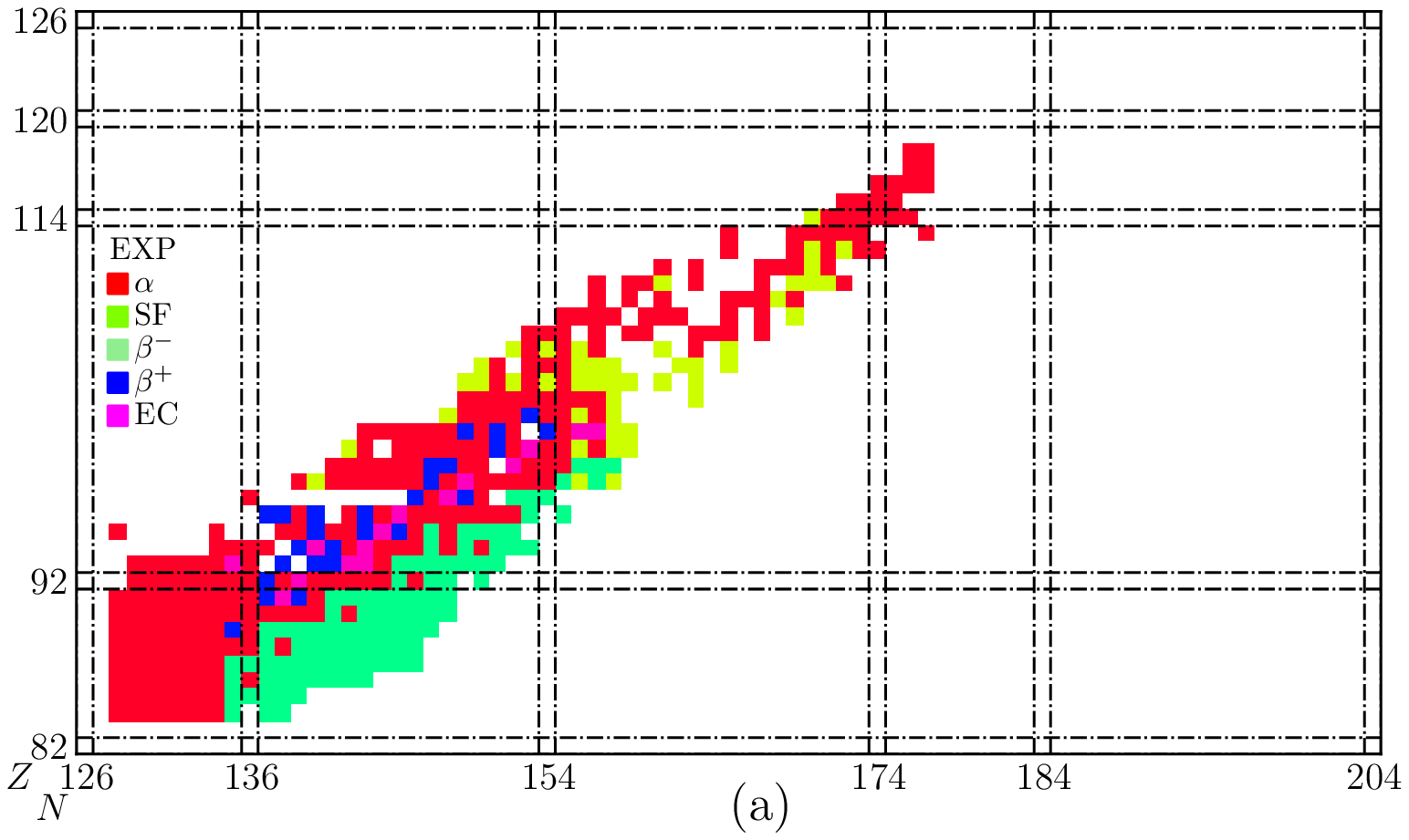}
\includegraphics[height=0.28\textwidth]{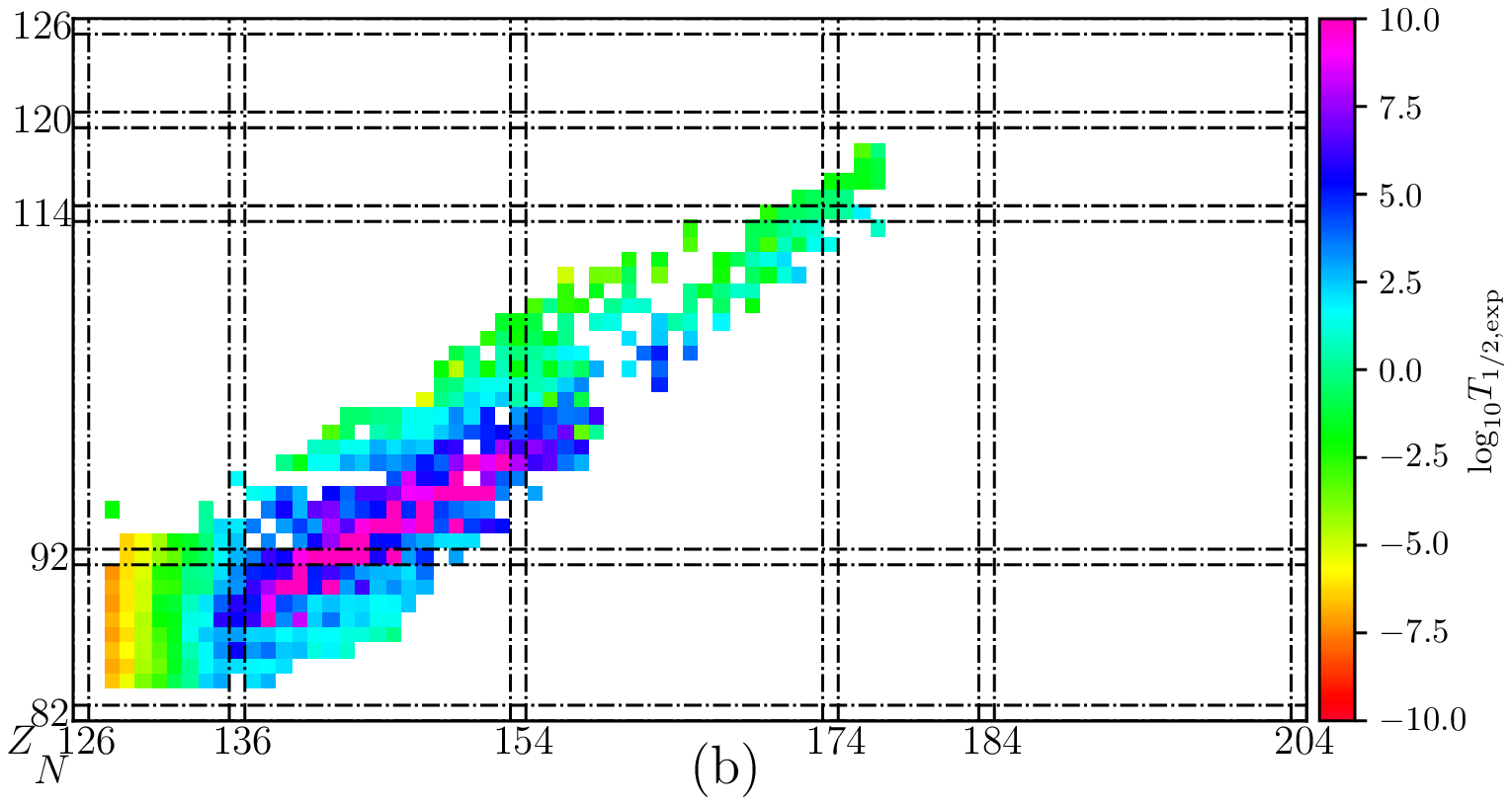}\\
\includegraphics[height=0.28\textwidth]{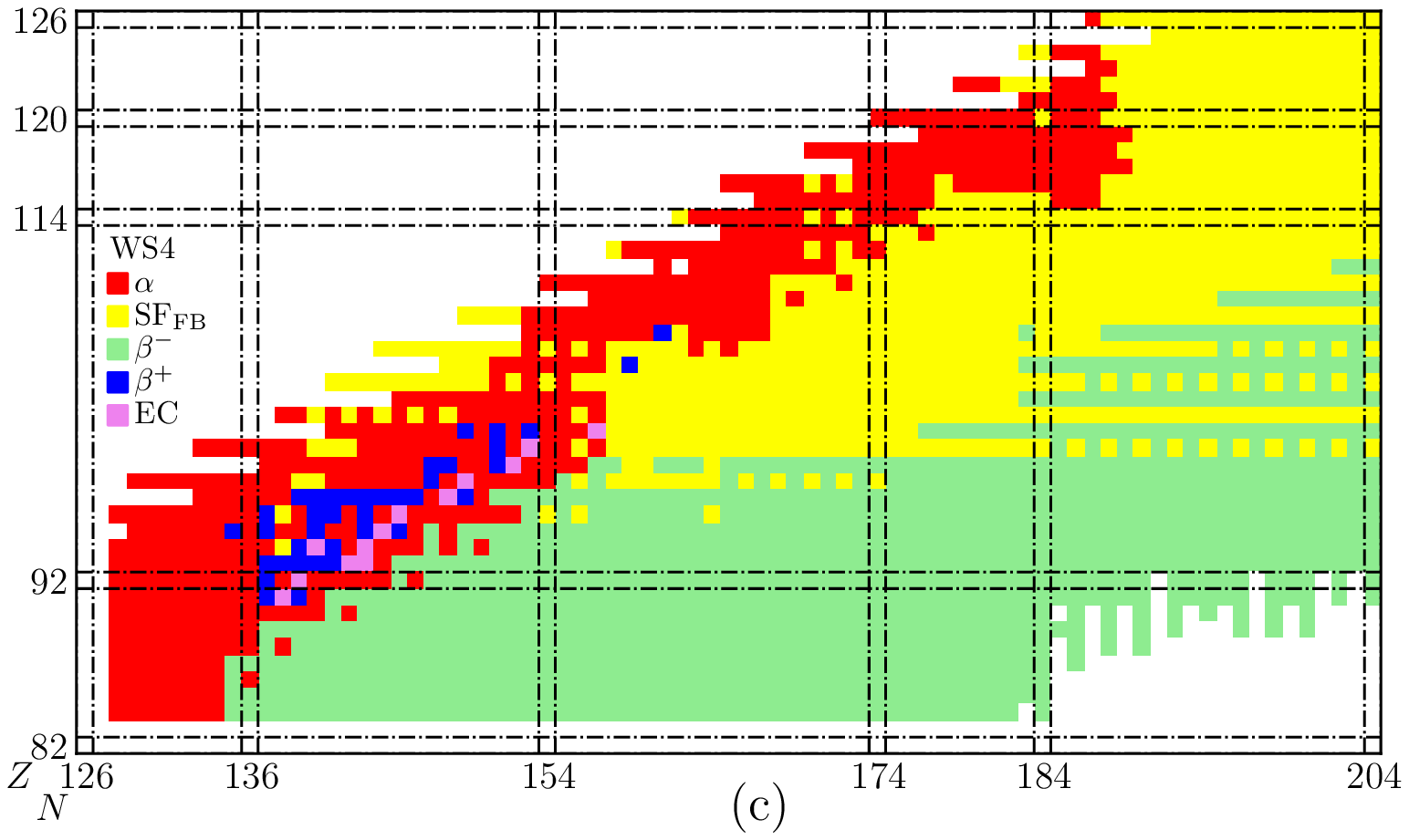}
\includegraphics[height=0.28\textwidth]{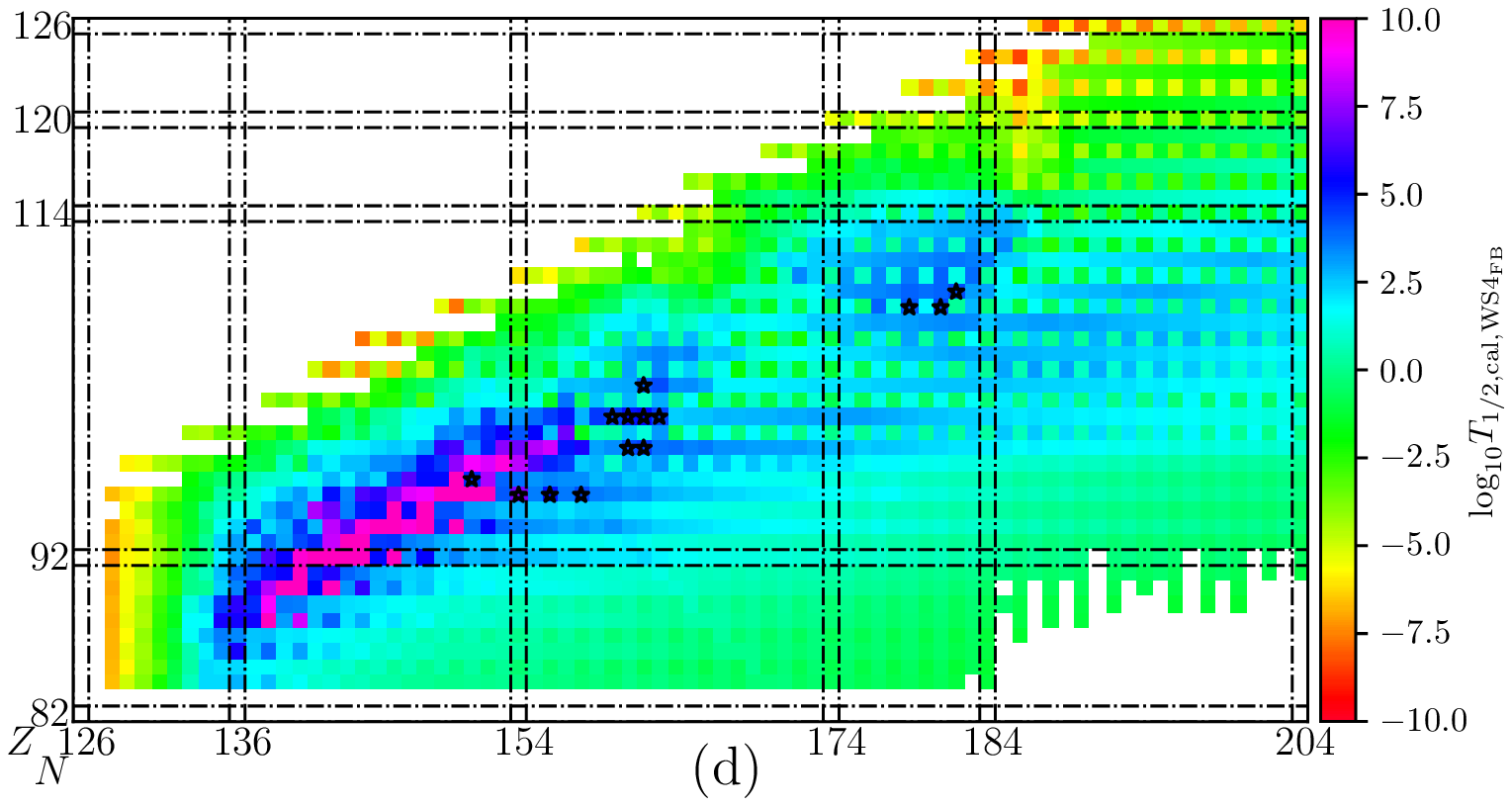}\\
\includegraphics[height=0.28\textwidth]{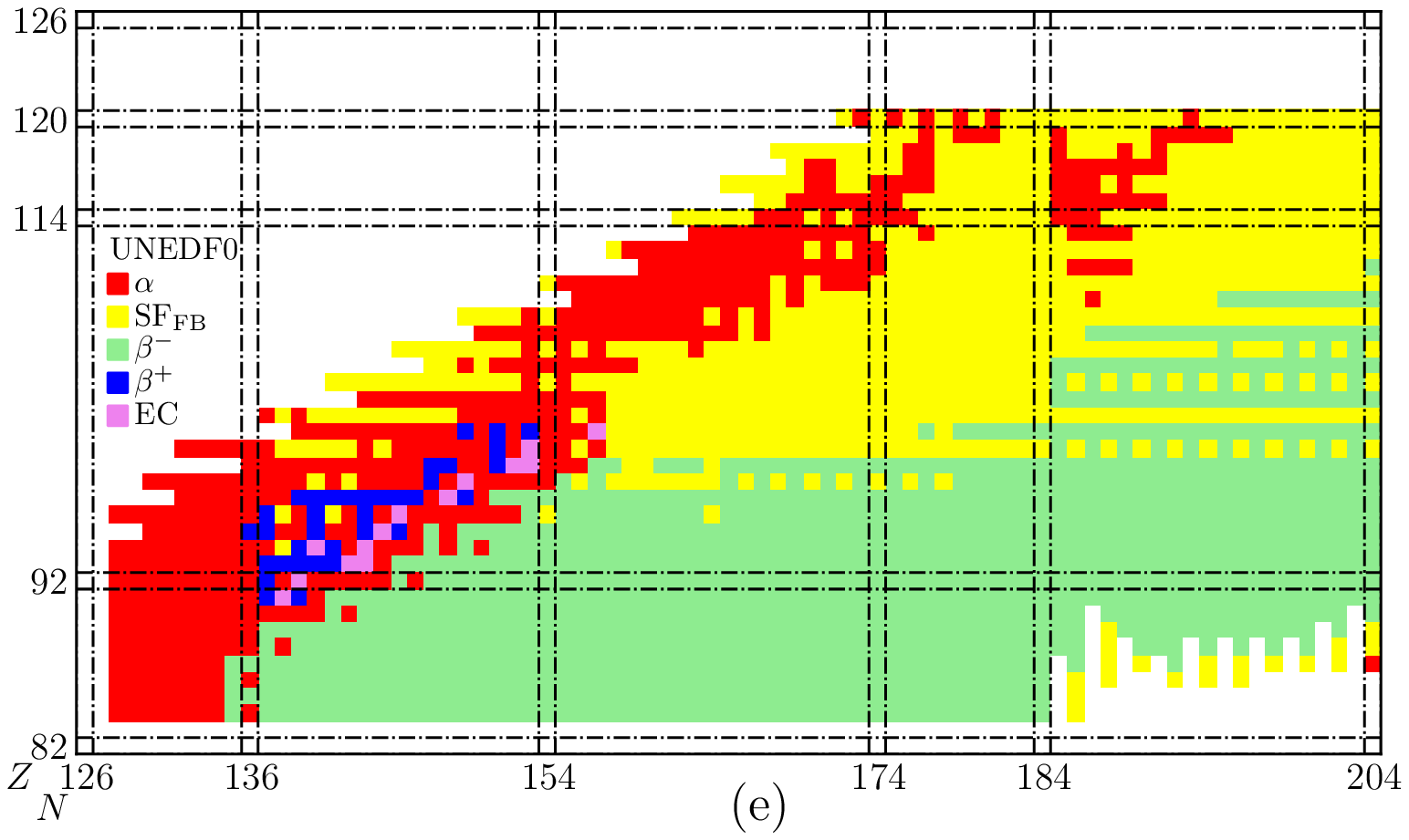}
\includegraphics[height=0.28\textwidth]{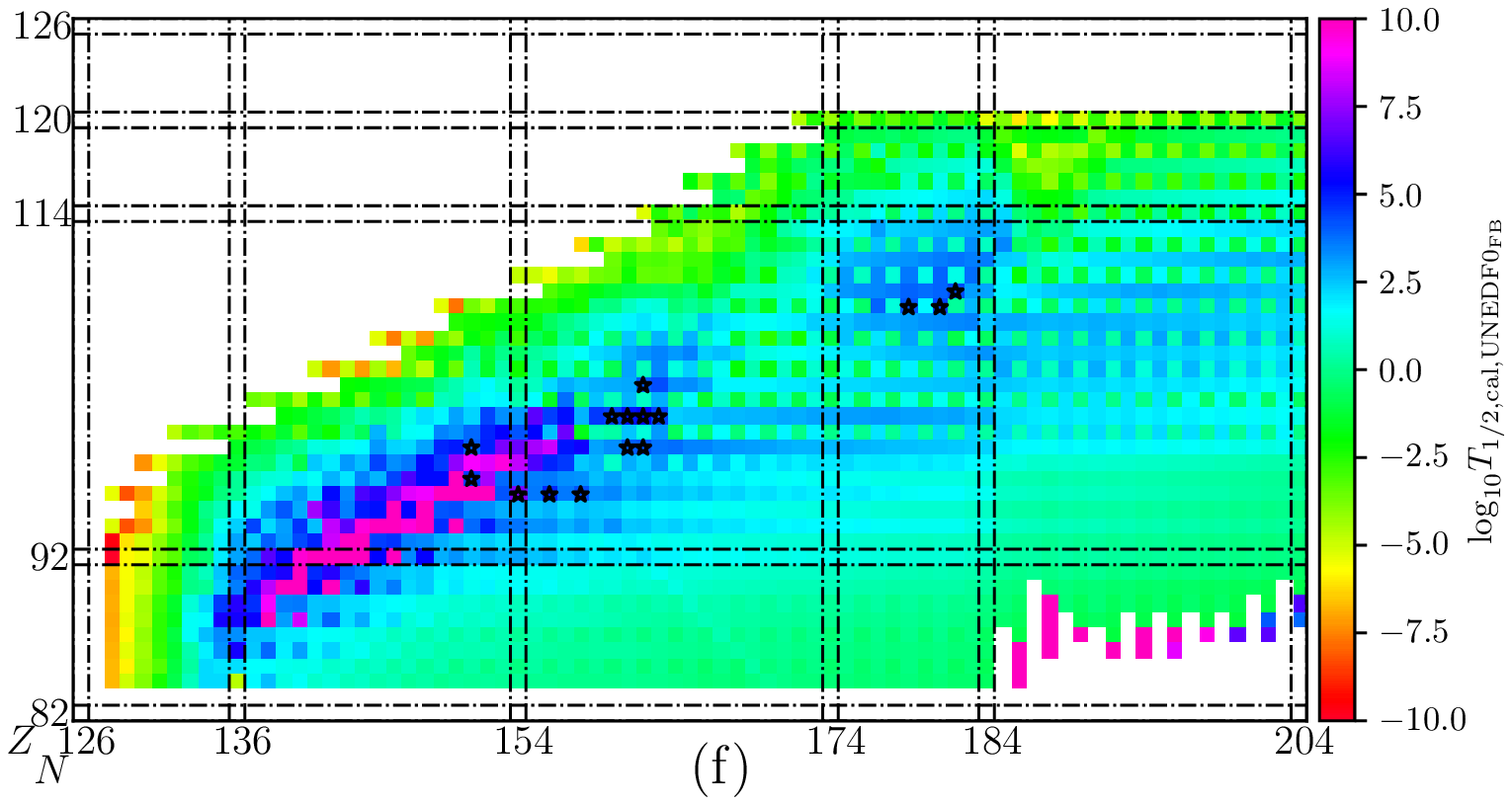}\\
\caption{\label{fig:lgT_heavy} The dominant decay mode (left panels) and the minimum partial half-lives (right panels) of $\alpha$ decay, $\beta^-$ decay, $\beta^+$ decay, EC and SF. (a-b) The experimental data in NUBASE2020. (c-f) The energies are calculated by WS4 and UNEDF0. Specifically, the FB is used to replace the decay energy to learn SF. Nuclides, of which the predicted partial half-life is longer than $10^4$ s, are marked by star. }
\end{center}
\end{figure*}

Totally, 445 nuclides with measured partial half-lives and branch ratios of these five decay modes are collected from NUBASE2020 \cite{kondev2021}. The dominant decay mode and the partial half-life of nuclides are drawn in FIG. \ref{fig:lgT_heavy}(a-b). A long-lived $\alpha$ decay valley from $^{226}_{88}$Ra$_{138}$ to $^{251}_{98}$Cf$_{153}$ lies between a narrow $\beta^+$/EC decay band and a neutron-rich $\beta^-$ region. Away from this valley, the half-life of nucleus decreases. The southwest direction is dominated by $\alpha$ decay while the southeast direction is occupied by $\beta^-$ decay. At the northwest direction, $\beta^+$ decay and EC compete with $\alpha$ decay and lose after $Z$ increases. At the northeast direction, $\alpha$ decay and SF compete with each other and a region extended from the $\alpha$ valley seems to be dominated by SF. Though the distribution of dominant decay mode has clear boundary, the minimum partial half-life is smooth.

There are 341 (104) nuclides are with known (unknown) corresponding reaction energies among all 445 considered nuclides. Those nuclides with unmeasured mass take that calculated by WS4 \cite{wang2014} and UNEDF0 \cite{kortelainen2010} for estimating the partial half-lives. The results are depicted in FIG. \ref{fig:lgT_heavy}(c-f). The calculation accords well with the experiment, as the dominant decay mode is correctly described for 431 and 427 (96.9\% and 96.0\%) nuclei when the RMSE of log$_{10}T_{1/2}$ of the dominant decay mode is 0.62 and 0.67, respectively.  Those nuclides, of which the dominant decay mode is inconsistently described, are generally with two competitive decay modes. For example, the $\alpha$ and SF branch ratios of $^{255}$Rf, $^{262}$Db, and $^{286}$Fl are around 50\%. Meanwhile, the liquid drop model trained by RF \cite{cai2022LDRF} is also applied to provide the energies, which presents consistent results and is not shown here.

The accuracy of energy is important to the half-life calculation. If calculated energies of WS4 and UNEDF0 replace all the experimental ones, the number with consistent decay mode comparing to the experiment reduces to 72.6\% and 64\%, and the RMSE of log$_{10}T_{1/2}$ also increases to 2.07 and 2.64. The difference of the results between using energies of the two models comes from the accuracy since the RMSE of mass of WS4 is about 0.3 MeV \cite{wang2014} while that of UNEDF0 is about 1.45 MeV \cite{kortelainen2010}. This also leads to difference during extrapolation. The consistency rate of dominant decay mode between using energies calculated by these two models decreases from 82.2\% to 66.2\%. More accurate and precise measurements of decay energy will aid the theoretical prediction. Besides, WS4 and UNEDF0 may lose prediction power after training through machine learning. If the WS4 and UNEDF0 binding energies are trained with features $Z$, $N$, $\delta$ and $P$, which well describe the residuals in Ref. \cite{niu2018}, though the description of energy improves, the consistency in the dominant decay mode decreases several percent, which is considerable compared with the rate 23.4\% of theoretical energies among all (104/445).

\begin{figure}[htbp]
\begin{center}
\includegraphics[width=0.5\textwidth]{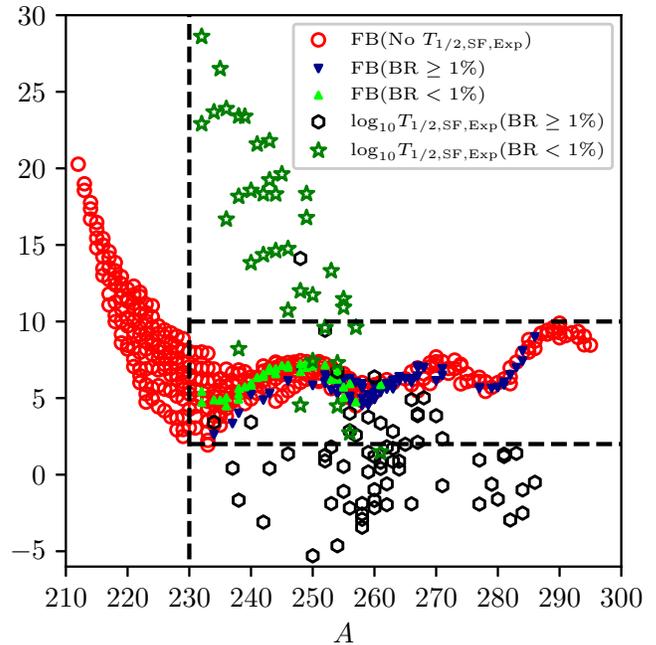}
\caption{\label{fig:Tsf_FB_epsilon-A} The evolution of $T_{1/2,\mathrm{SF}}$ and FB  along the mass number. The datasets are divided according to whether the corresponding $T_{1/2,\mathrm{SF}}$ is measured and whether the branch ratio (BR) of SF is less than 1\%.}
\end{center}
\end{figure}

The SF is important for investigating the half-lives of superheavy nuclei. As shown by FIG. \ref{fig:lgT_heavy}(c, e), the dominant decay mode of the unknown nuclides is determined through the competition between SF, $\alpha$ decay, and $\beta^-$ decay. The major competition is between SF and $\beta^-$ decay for neutron-rich nuclides, while it is between SF and $\alpha$ decay for neutron-deficient ones. The existing experimental data show a long-lived $\alpha$ decay region from $^{226}_{88}\rm{Ra}_{138}$ to $^{251}_{98}\rm{Cf}_{153}$, lying between the $\beta^+$ and $\beta^-$ decay regions, and being ended by the SF. The present models correctly describe such phenomenon. Along the long-lived region, and after $N$ exceeds 154, a blue band is shown in FIG. \ref{fig:lgT_heavy}(d, f), which indicates half-lives of about $10^2\sim10^7$ s. At the southwest corner of $Z=114$ and $N=184$, nuclides in a circle are with longer half-life than those surrounded. This is because the fission barrier is high in this region and lead to longer $T_{1/2, \mathrm{SF}}$. FIG. \ref{fig:Tsf_FB_epsilon-A} compares the evolution of FB and measured $T_{\mathrm{1/2, SF}}$ along the mass number. The FB decreases with $A$ before $A=230$, then behaves like a sinusoidal wave oscillating between 2 and 10 MeV. It seems that there exists a FB threshold, below which could the nuclide fissure spontaneously. Nuclides with rather long $T_{1/2, \mathrm{SF}}$ are generally with small SF branch ratio.  Besides, the FB of nuclides with SF branch ratio less than 1\% is mostly higher than those with SF branch ratio greater than 1\%, which implies that higher the FB, weaker the SF. However, if one concentrates only on nuclides with SF branch ratio less than 1\% or greater than 1\%,  the correspondence between FB and $T_{1/2, \mathrm{SF}}$ becomes much more complex.

\begin{figure*}[htbp]
\begin{center}
\includegraphics[width=0.325\textwidth]{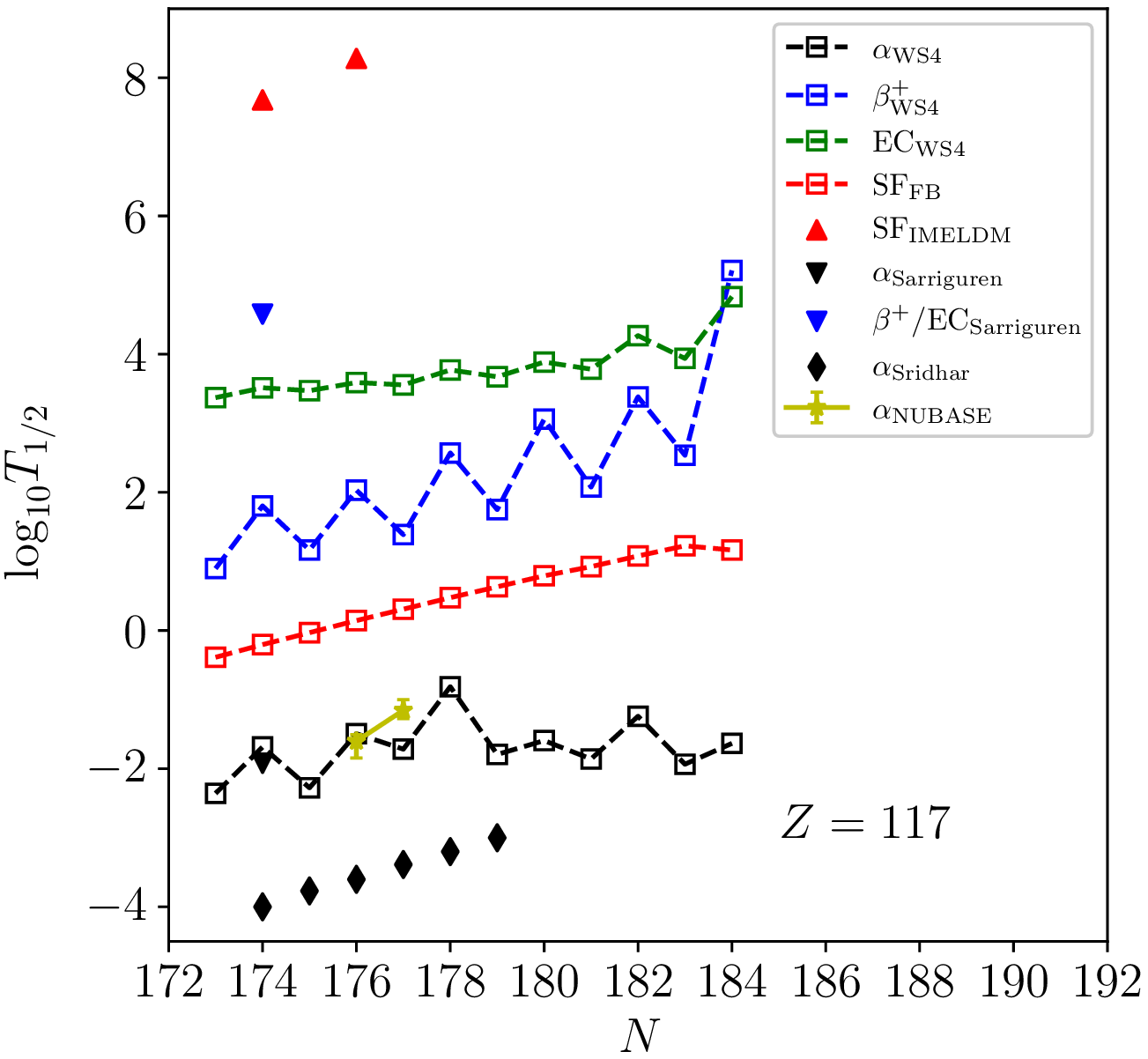}
\includegraphics[width=0.325\textwidth]{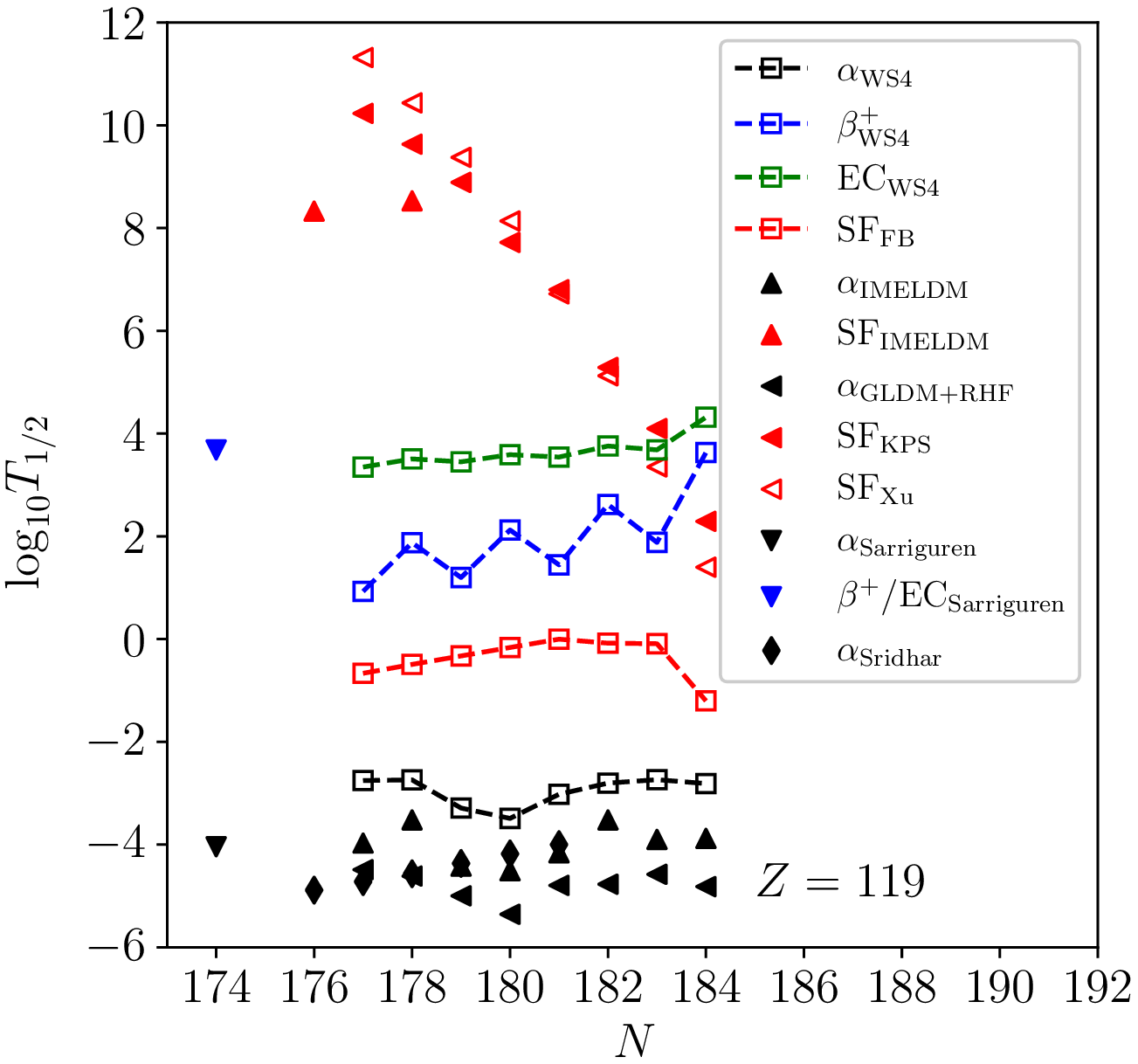}
\includegraphics[width=0.325\textwidth]{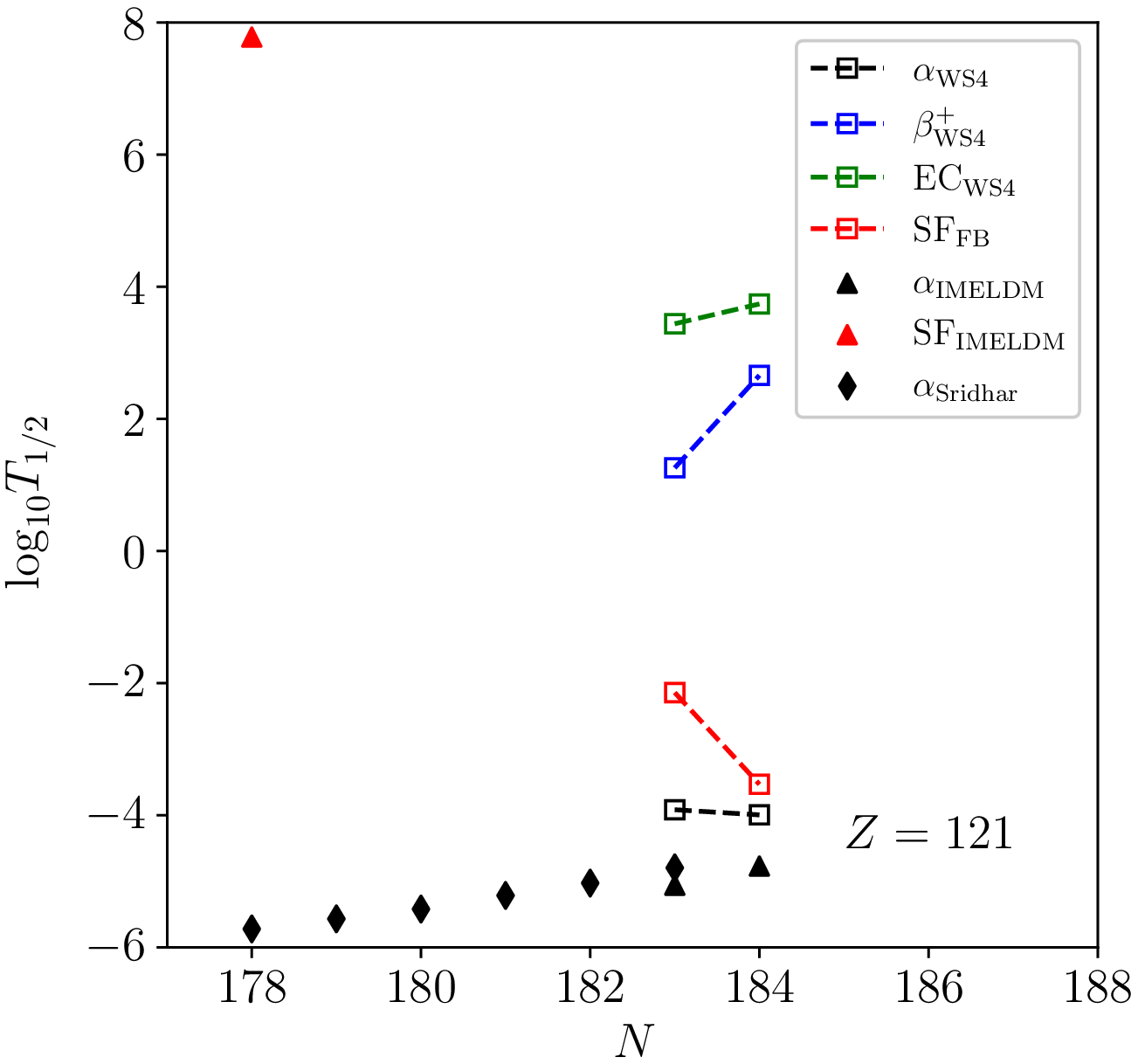}\\

\includegraphics[width=0.325\textwidth]{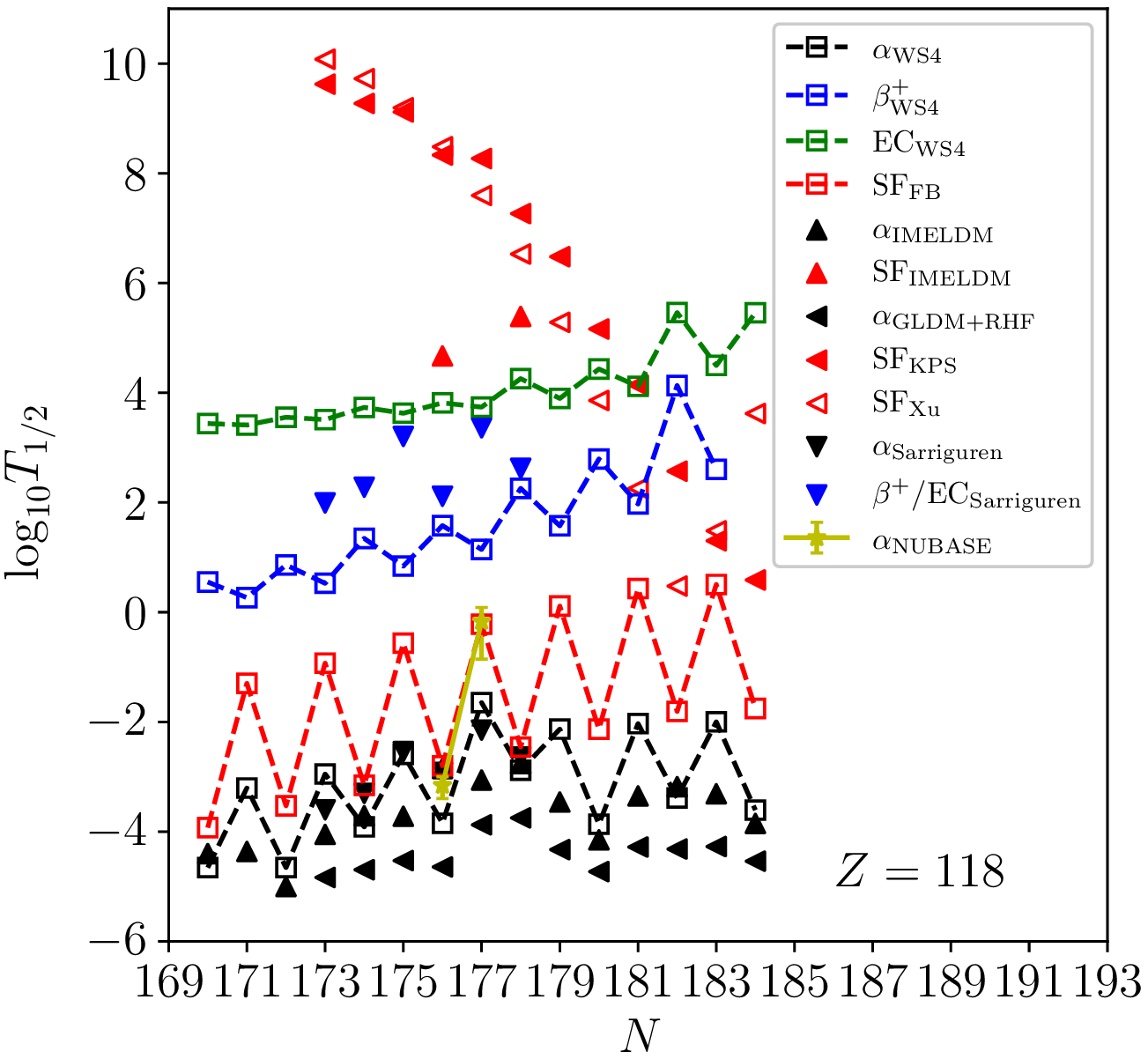}
\includegraphics[width=0.325\textwidth]{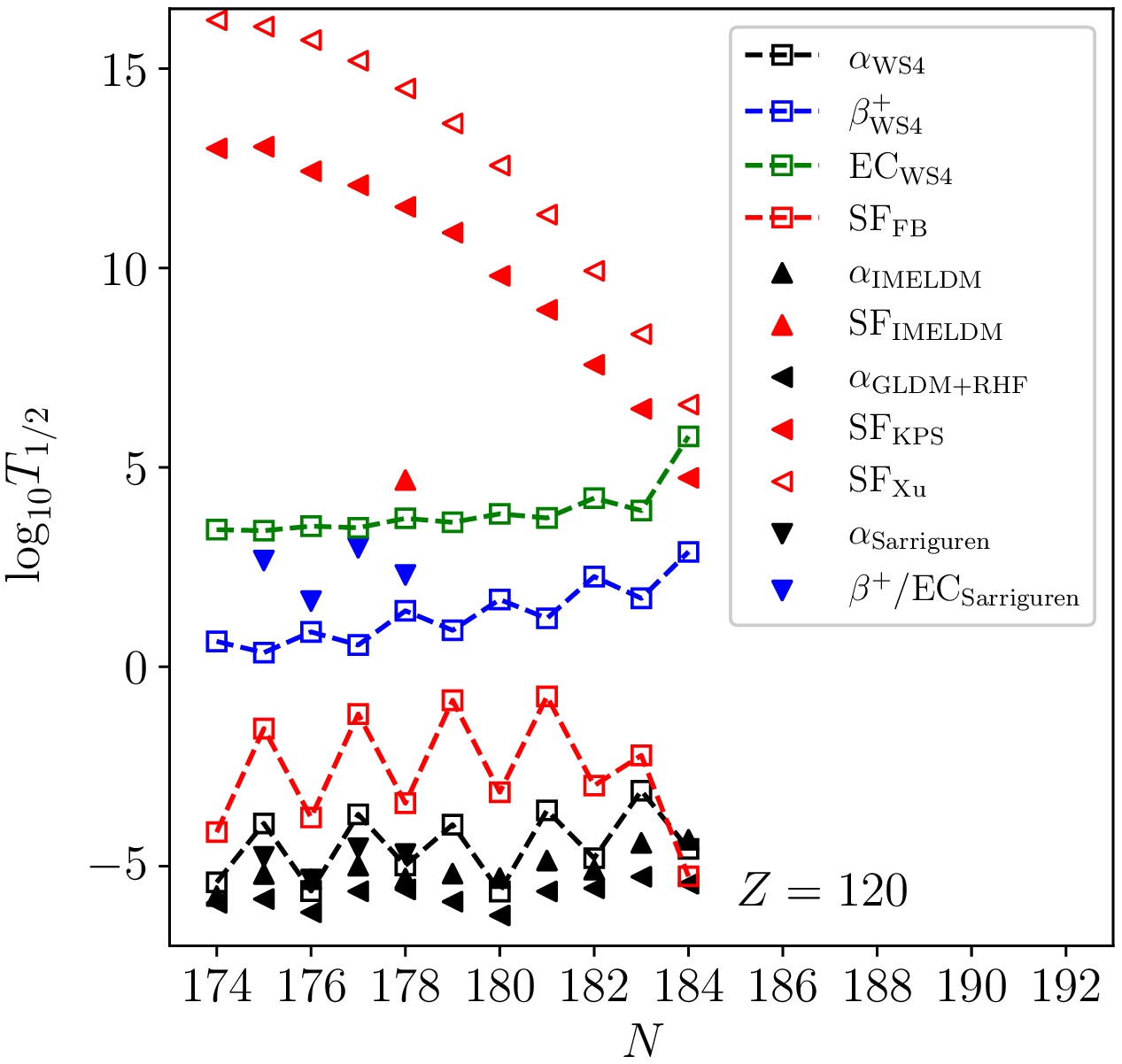}
\includegraphics[width=0.325\textwidth]{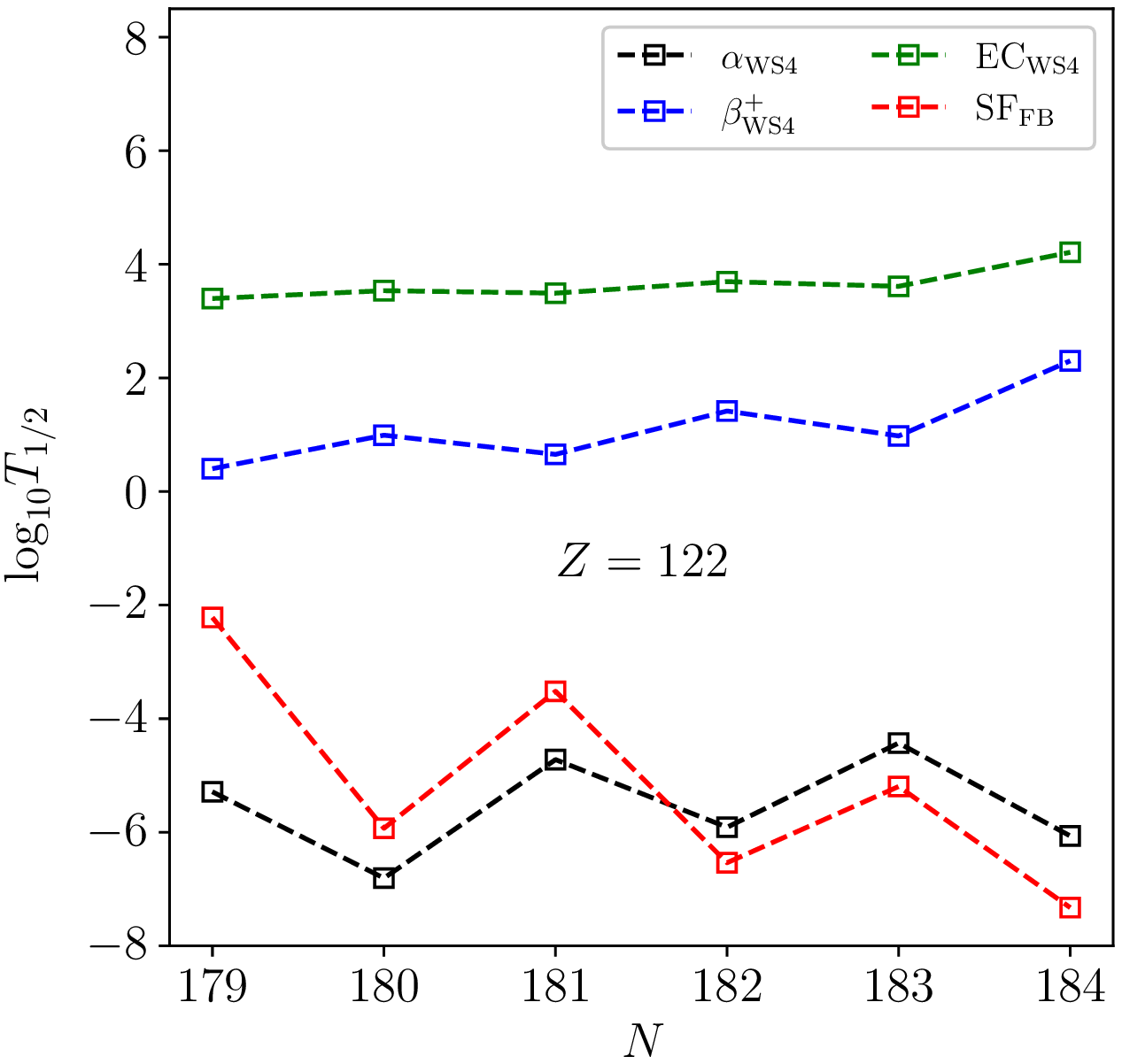}
\caption{\label{fig:lgT_new_isotopes}Comparison of partial half-lives of isotopes with $Z=117-122$. IMELDM is extracted from Ref. \cite{cui2022}, GLDM+RHF, KPS and Xu are extracted from \cite{he2022}, Sarriguren is extracted from Ref. \cite{sarriguren2021}, Sridhar is extracted from Ref. \cite{sridhar2018}.}
\end{center}
\end{figure*}

Nuclides with partial half-life predicted to be longer than $10^4$ s are marked by star in FIG. \ref{fig:lgT_heavy}(d, f), which suggests $^{250,252,254}\rm{Cm}$, $^{260, 261}\rm{Es}$, $^{261\sim264}\rm{Md}$, and $^{265}\rm{Lr}$ for future measurement. It is a coincidence that no experimental value of the half-life of $^{250}$Cm is suggested in NUBASE2020 and thus extrapolated in the present work. In NNDC, SF is shown to be its dominant decay mode, and the half-life is recommended to be 8300 years, which is rather long. Though the calculation of the present work underestimates the NNDC value, the long half-life property and the dominant decay mode are reproduced. Besides, the upper limit of the half-life of $^{252}$Cm is 2 days, proposed in 1966 by Ref. \cite{combined1966} and not updated, while the present work estimates a value of 1.43 days. No experimental half-lives are published for $^{260, 261}$Es, $^{261\sim264}$Md, and $^{265}$Lr. But their nearby isotopes are with long half-lives, e.g., $^{257}$Es (7.7 days), $^{260}$Md (31.8 days), $^{259}$Md (1.6 hours), $^{258}$Md (51.5 days), $^{257}$Md (5.52 hours) and $^{266}$Lr (11 hours). Moreover, these Es, Md, and Lr isotopes locate at the extension of the narrow long-lived region from $^{226}$Ra to $^{251}$Cf, which makes it convincing that these Es, Md, and Lr isotopes are candidates with long partial half-lives. More measurement is also suggested since, for example, the more than 50 years of un-updated datum of $^{252}$Cm.

Comparing all possible decay channels is limited by the accurate description of each channel and the observed data. We should note that the mechanism of SF is still not fully understood. The effect of the quadrupole deformation parameter ($\varepsilon_2$) \cite{moller2016} on the half-life estimation is also investigated. If one replaces FB by $\varepsilon_2$ during training the RF, more nuclides in $N>184$ are calculated to be dominated by $\alpha$ and $\beta^-$ decay. The $\alpha$ decay leads to shorter half-life. Besides, the relatively long-lived circle at the southwest of $Z=114, N=184$ is no longer locally. The blue region is much more extended and more long-lived candidates are marked by star. This is because the deformation decreases, which compensates the Coulomb repulsion, that is increased by $Z$. Furthermore, the FB combines contribution of multipole deformations and thus presents stronger quantum effect in FIG. \ref{fig:lgT_heavy}(d, f) than $\varepsilon_2$.

The extrapolation stops at the single neutron (proton) and two-neutron (two-proton) drip lines. Dataset of UNEDF0 stops at $Z=120$.  From the existent region to the neutron-deficient side, $\alpha$ decay and SF are predicted to compete with each other. On the neutron-rich side, calculations predict $\beta^-$ decay as the dominant mode, while SF competes in specific nuclides. Up to date, results of most theoretical calculations of partial half-lives  \cite{sarriguren2021, cui2022, he2022, sarriguren2022, sexena2021, xu2013, bao2015, liu2017, soylu2019, sridhar2018, sahoo2020} support that the $\alpha$ decay is the dominant decay mode for new elements at $N\leqslant 184$. But those calculated $T_{1/2, \rm{SF}}$ diverge when it is far from shells. In FIG. \ref{fig:lgT_new_isotopes}, we compare the  partial half-lives of isotopes with $Z=117-122$ predicted in the present work and the corresponding results of Refs. \cite{sarriguren2021, cui2022, he2022}. Even though the partial half-lives of $\beta^+$ decay and EC in the present work are not globally longer than that in Ref. \cite{sarriguren2021}, they are still about five orders of magnitude greater than that of $\alpha$ decay in this region, which does not change the dominant decay mode.

\begin{figure}[htbp]
\begin{center}
\includegraphics[width=0.5\textwidth]{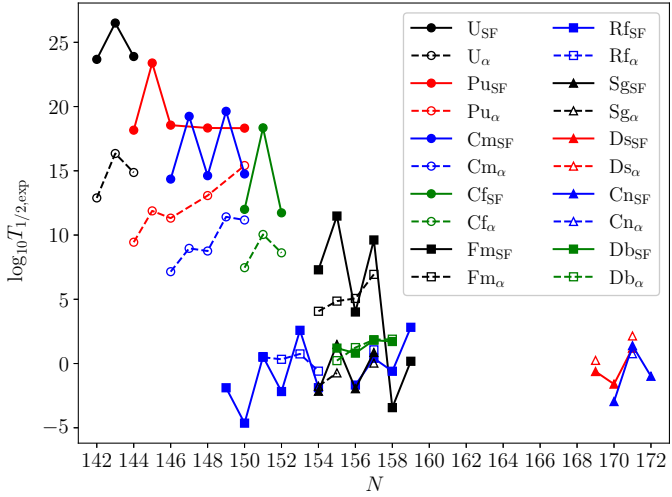}
\caption{\label{fig:lgTsf_OES} The odd-even staggering of $T_\mathrm{1/2, SF}$ of U, Pu, Cm, Cf, Fm, Rf, Sg, Ds, Cn and Db isotopes and comparing with $T_{1/2, \alpha}$.}
\end{center}
\end{figure}

$T_{1/2, \alpha}$ predicted in the present work are longer than the results of Refs. \cite{sarriguren2021, cui2022, he2022}, which does not change the dominant decay mode of odd-$Z$ isotopes but enhances the competition of SF in even-$Z$ isotopes. Furthermore, the prediction of the present work shows strong odd-even staggering of $T_{1/2, \rm{SF}}$ of even-$Z$ isotopes, i.e., $T_{1/2, \rm{SF}}$ of even-even nuclei is several times of magnitude shorter than its two isotopic neighbors, which differs from the weak or not predicted odd-even staggering effect of other SF models in FIG. \ref{fig:lgT_new_isotopes}. In fact, all measured $T_{1/2, \rm{SF}}$ of even-$Z$ isotopes show such odd-even staggering. FIG. \ref{fig:lgTsf_OES} draws $T_{1/2, \mathrm{SF}}$ and $T_{1/2, \alpha}$ of isotopes with $Z\geqslant92$. When $Z$ is small, for example in U, Pu, Cm and Cf isotopes, SF is not competitive to $\alpha$ decay because the Coulomb repulsion is not enough strong. But when $Z$ is large, this odd-even staggering makes SF competitive with $\alpha$ decay in these even-even nuclides. The $\alpha$ decay is thus suggested to be a key signal detected for $Z=119$ and $121$ isotopes, while the SF should also be taken into account for even-$N$ isotopes of $Z=120$ and $122$. Note that the odd-even staggering also exists in the odd-$Z$ isotopes. It can only be verified by $^{260\sim263}$Db because the data are rare. This is why the odd-even staggering of odd-$Z$ isotopes is not predicted in the present work. The DNS model predicted $\sigma_{\rm{ER}}$ of hundreds of fb for the 3$n$ or 2$n$ channels producing $^{293}119_{174}$ on $^{243}$Am target \cite{li2018}, which can be examined on the new facility CAFE2 and SHANS2 in Lanzhou \cite{sheng2021}. Considering the odd-even effect of partial half-lives, candidates of nuclide for new superheavy elements still need analysis through the production cross section and the partial half-life.

\begin{figure}[htbp]
\begin{center}
\includegraphics[width=0.45\textwidth]{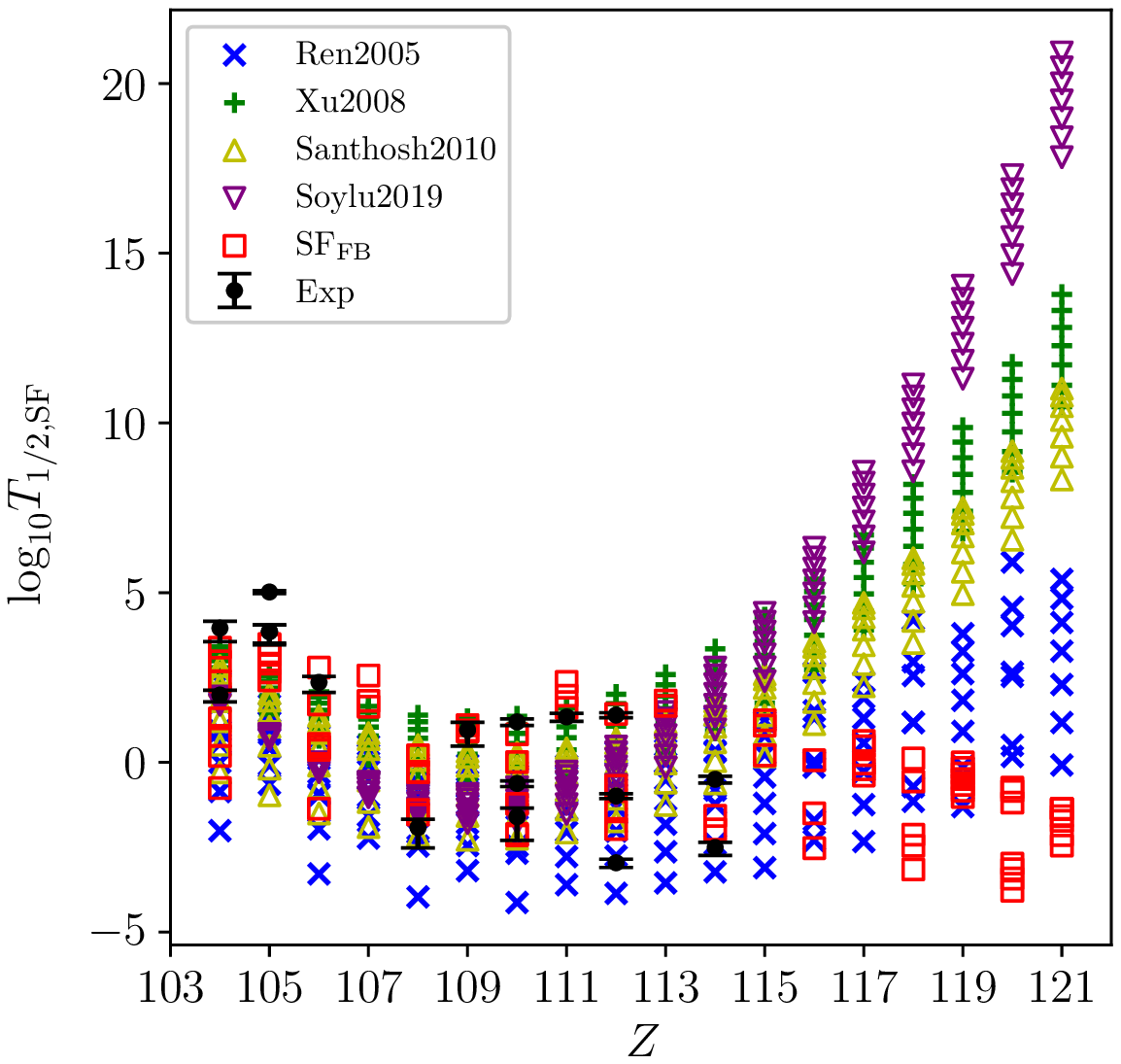}
\includegraphics[width=0.45\textwidth]{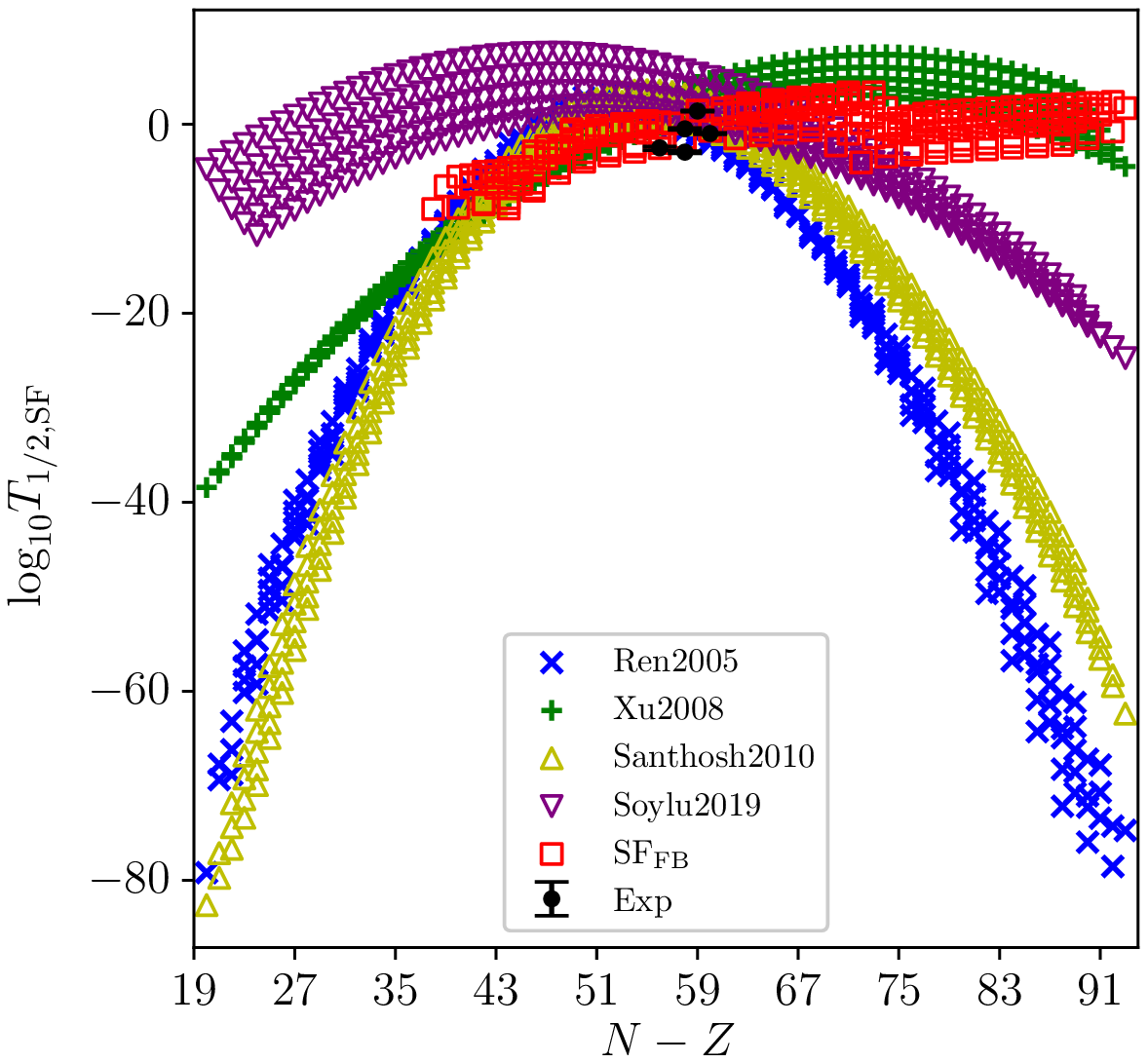}
\caption{\label{fig:lgTsf_compare} The comparison of $T_{1/2, \rm{SF}}$ calculated by formulas of Ren2005 \cite{ren2005}, Xu2008 \cite{xu2008}, Santhosh2010 \cite{santhosh2010}, Soylu2019 \cite{soylu2019} and the present work ($\mathrm{SF_{FB}}$). The extrapolation in the upper panel is constrained by $104\leqslant Z\leqslant121$ and  $56\leqslant N-Z\leqslant62$, the bottom panel is constrained by $Z=112-116$.}
\end{center}
\end{figure}

There are several formulas of the SF with more parameters \cite{ren2005, xu2008, santhosh2010, soylu2019}, but the extrapolation diverges, sometimes to tens of orders of magnitude, as drawn in Figure \ref{fig:lgTsf_compare}, because of the higher order terms. There is no hint of such divergence of $T_{\rm{1/2, SF}}$ from experiments. If the SF is not considered in the extrapolation because of its possible large uncertainty, a long-lived region is predicted along the boundary of competition between the $\alpha$ and $\beta^-$ decays. Particularly, results from UNEDF0 show a new long-lived region just around $Z=114$ and $N=184$, which is also mainly contributed by the $\alpha$ decay. Thus, the SF is key for investigating the superheavy stable island but still not well understood.

\section{Conclusion\label{sec:conclusion}}

In summary, the decay modes of superheavy nuclei are investigated through the Random Forest algorithm. The partial half-lives of $\alpha$ decay, $\beta^-$ decay, $\beta^+$ decay, EC, and SF are studied and compared with each other. The dominance of $\alpha$ decay in the neutron-deficient region is more convinced. $\beta^-$ decay is predicted to be dominant in the neutron-rich region. SF contributes to a long-lived circle at the southwest corner of $Z=114$ and $N=184$. More accurate and precise measurement on nuclear mass and decay energy can improve the prediction of decay mode. After correcting the divergence of up-to-date SF formulas, the odd-even effect of SF is found in even-$Z$ nuclides, which induces possible competition between SF and $\alpha$ decay in even-even nuclides. The $\alpha$ decay is thus suggested to be key probe of isotopes with $Z=119$ and $121$, while the competition of  SF should be taken into account in even-even isotopes with $Z=120$ and $122$. 

$^{250, 252, 254}\rm{Cm}$, $^{260, 261}\rm{Es}$, $^{261\sim264}\rm{Md}$, and $^{265}\rm{Lr}$ with half-life predicted to be longer than $10^4$ s are suggested for future measurement. The SF influenced by the fission barrier and the Coulomb repulsion leads to a long-lived region during extrapolation.  The present results indicate that study on SF, especially beyond $^{286}$Fl, which is the presently heaviest nuclide with significant SF branch ratio, would be of key importance for further studies to be performed on new facilities, such as CAFE2 and SHANS2 in Lanzhou.

\begin{acknowledgments}
The authors acknowledge useful discussion with Professors Zaiguo Gan, Zhongzhou Ren, and Zhiyuan Zhang. This work has been supported by the Guangdong Major Project of Basic and Applied Basic Research under Grant No. 2021B0301030006, and the computational resources from SYSU and National Supercomputer Center in Guangzhou. 

\end{acknowledgments}

\bibliography{SHN_Life_Mode.bib}

\end{document}